\documentstyle[11pt]{article}
\begin{document}
\parindent 0.7cm

\begin{titlepage}
\centerline{\large\bf
       Estimate of Wolfenstein's Parameters $\rho$ and $\eta$}
\centerline{\large\bf
           Based on a Geometry viewpoint to the Weak CP Phase }

\vspace{1cm}

\centerline{ Yong Liu }
\vspace{0.5cm}
{\small
\centerline{\bf  Laboratory of Numeric Study for Heliospheric Physics}
\centerline{\bf  Chinese Academy of Sciences,
                 P. O. Box 8701, Beijing 100080, P.R.China}
}
\vspace{2cm}

\centerline{\bf Abstract}
\vspace{0.3cm}

Based on a geometric postulation on the weak $CP$ phase in
Cabibbo-Kobayashi-Maskawa (CKM) matrix, a positive $\rho$ is
asserted. Besides, $0.18 <\eta< 0.54$ and $0.048<\rho< 0.140$ are
permitted by the present data. The corresponding geometric
constraint on Wolfenstein's parameters is also worked out. We find
that, according to the geometry viewpoint, $\eta$ and $\rho$
satisfy an approximate linear relation. These results can be put to
the more precisely tests in near future.
\\\\
PACS number(s): 12.10.Ck, 13.25.+m, 11.30.Er

\vspace{5cm}
\noindent
Email address: yongliu@ns.lhp.ac.cn
\end{titlepage}

\vspace{1cm}

Quark mixing and $CP$ violation is one of the most interesting and
important problem in weak interaction [1-4]. In the Standard Model,
They are described by the unitary Cabibbo-Kabayashi-Maskawa (CKM)
matrix, which takes the following form [5-7]
\begin{equation}
V_{KM}= \left (
\begin{array}{ccc}
   c_{12} c_{13} & s_{12} c_{13}& s_{13} e^{-i \delta_{13}} \\
   -s_{12} c_{23}-c_{12} s_{23} s_{13} e^{i \delta_{13}} &
   c_{12} c_{23}-s_{12} s_{23} s_{13} e^{i \delta_{13}}    &
   s_{23} c_{13}\\
   s_{12} s_{23}-c_{12} c_{23} s_{13} e^{i \delta_{13}}  &
   -c_{12} s_{23}-s_{12} c_{23} s_{13} e^{i \delta_{13}} &
   c_{23} c_{13}
\end{array}
\right )
\end{equation}
in the standard parametrization. Here, the standard notations
$c_{ij}=\cos\theta_{ij}$ and $s_{ij}=\sin\theta_{ij}$ for the
"generation" labels $i,j=1,2,3$ being used. The real angles
$\theta_{12},\; \theta_{23}$ and $\theta_{13}$ can all be made to
lie in the first quadrant while the phase $\delta_{13}$ lies in the
range $0<\delta_{13}<2 \pi$. In following, we will fix the three
angles $\theta_{12}, \theta_{23}$ and $\theta_{13}$ in the first
quadrant.

To make it be convenient to use the CKM matrix in the concrete
calculations, Wolfenstein parametrized it as [8]
\begin{equation}
V_{W}= \left (
\begin{array}{ccc}
   1-\frac{1}{2}\lambda^2 & \lambda & A\lambda^3(\rho-
   i\eta+i\eta\frac{1}{2}\lambda^2) \\
   -\lambda & 1-\frac{1}{2}\lambda^2-i\eta A^2 \lambda^4 &
   A\lambda^2(1+i\eta\lambda^2)\\
   A\lambda^3(1-\rho-i\eta) & -A\lambda^2 & 1
\end{array}
\right ).
\end{equation}

Actually, one can take different parametrization [7][9-13] in
different cases. They are only for the convenience in discussing
the different questions, but the physics does not change when
adopting various parametrizations.

According to Buras etc., there is a very nice corresponding
relation between Wolfenstein's parameters and the ones in the
standard parametrization. It reads [14]
\begin{equation}
s_{12}=\lambda, \;\;\;\;\;\; s_{23}=A\lambda^2, \;\;\;\;\;\;
s_{13}e^{-i \delta_{13}}=A\lambda^3 (\rho-i\eta).
\end{equation}
So,
\begin{equation} s_{13}=A\lambda^3 \sqrt{\rho^2+\eta^2},
\;\;\;\;\;\;
\sin\delta_{13}=\frac{\eta}{\sqrt{\rho^2+\eta^2}}
\end{equation}
and consequently,
\begin{equation}
\rho=\frac{s_{13}}{s_{12}s_{23}}\cos \delta_{13}, \;\;\;\;\;\;
\eta=\frac{s_{13}}{s_{12}s_{23}}\sin \delta_{13}.
\end{equation}

In Eq.(2), $\lambda$ and $A$ are the two better known parameters.
But, due to the uncertainty of hadronic matrix elements and other
reasons, it is difficult to extract more information about $\rho$
and $\eta$ from experimental results. Up to now, we still know
little about them. More than ten years ago, Wolfenstein estimated
that the upper limit on $\eta$ is about $0.1$ [8], but the recent
analysis indicate that, $\rho$ and $\eta$ are about [9][15]
\begin{equation}
-0.15< \rho <0.35, \;\;\;\;\;\;\;\;\;\; 0.20< \eta<0.45.
\end{equation}

The central purpose of this work is to give limits on the ranges of
$\rho$ and $\eta$ according to our geometric postulation on the
weak $CP$ phase.

In Ref. [16], we have found that, the weak $CP$ phase and the other
three mixing angles in CKM matrix satisfy the following geometry
relation
\begin{equation}
\sin\delta_{13}=\frac{ (1+s_{12}+s_{23}+s_{13})
                       \sqrt{1-s_{12}^2-s_{23}^2-s_{13}^2+
                       2 s_{12} s_{23} s_{13}} }{(1+
                       s_{12}) (1+s_{23}) (1+s_{13})}.
\end{equation}
Here, we have taken $\delta_{13}$ as certain geometry phase. In
fact, Eq.(7) means that, $\delta_{13}$ is the solid angle enclosed
by $(\pi/2-\theta_{12})$, $(\pi/2-\theta_{13})$ and
$(\pi/2-\theta_{23})$, or the area to which the solid angle
corresponds on a unit sphere. Obviously, to make
$(\pi/2-\theta_{12})$, $(\pi/2-\theta_{13})$ and
$(\pi/2-\theta_{23})$ enclose a solid angle, the condition
\begin{equation}
(\pi/2-\theta_{ij})+(\pi/2-\theta_{jk})>(\pi/2-\theta_{ki})
\;\;\;\; (i\neq j \neq k \neq i.
\;\; i,j,k=1,2,3.\;\; \theta_{ij}=\theta_{ji})
\end{equation}
must be satisfied.

Now that the four angles in CKM matrix are not independent, the
four Wolfenstein's parameters $A, \; \lambda,\; \rho$ and $\eta$
must also be not independent.

Substituting Eqs.(3-4) into Eq.(7), it is easy to achieve
\begin{equation}
\frac{\eta}{\sqrt{\rho^2+\eta^2}}=1-\frac{\lambda^2}{2}-A \lambda^3
+\lambda^4 (-\frac{1}{8}+A-\frac{A^2}{2}-A \sqrt{\rho^2+\eta^2}).
\end{equation}
This is just the geometry constraint on Wolfenstein's parameters
when approximate to the fourth order of $\lambda$.

It should be noted that, $\rho$ presents as its square in the
equation, hence, its sign is ambiguous. In fact, this kind of
vagueness is inavoidable. Because the observables are the
combinations of $\rho$ and $\eta$ rather than themselves, in which
$\rho$ always presents as its square, so, it is very difficult to
specify the sign of $\rho$ from the experimental data.

Here, we would like to give a comment on the sign of $\rho$.
According to our geometric postulation, as discussed in Ref. [16],
when fixing all the three mixing angles in the first quadrant,
$\delta_{13}$ must be in the first quadrant, or at most, the fourth
quadrant is permitted. In either case, Eq.(5) implies a positive
$\rho$.

In following, let us investigate carefully the permitted ranges of
$\rho$ and $\eta$ by present data. If we start out from Eq.(9)
directly, and take [7][17]
$$
\lambda=0.2196\pm0.0023 \;\;\;\;\;\; A=0.819\pm 0.035
$$
as inputs, then we can obtain the dependence of $\eta$ on $\rho$.
The result is shown in Fig.(1). It can be seen from the figure
that, $\eta$ and $\rho$ satisfy an approximate linear relation.

We can also begin with Eq.(5). But, we should know the three mixing
angles firstly. This can be arrived by use of three of the CKM
matrix elements such as $V_{ud}$, $V_{ub}$ and $V_{tb}$. In Ref.
[16], we have found that, the whole matrix can be reconstructed
very well based only on three of the elements and Eq.(7). Once the
three mixing angles are determined, we can extract the dependence
of $\eta$ on $\rho$ again from Eq.(5). We take the relevant inputs
from the data book [7]
$$
V_{ud}=0.9745\sim 0.9760, \;\;\;\;\;\; V_{ub}=0.0018\sim 0.0045,
\;\;\;\;\;\; V_{tb}=0.9991 \sim 0.9993.
$$
The numerical result is also shown in Fig.(1). We find it is just a
little part of that drawn from Eq.(9).

Now, we can read from the figure that, when all the three inputs
$V_{ud},\; V_{ub}$ and $V_{tb}$ are taken at $95\% \;\; C.L.$, we
obtain the outputs
\begin{equation}
0.048<\rho<0.140, \;\;\;\;\;\;\;\;\;\; 0.18<\eta<0.54.
\end{equation}

Comparing with Eq.(6), the range for $\rho$ has been narrowed down
while the limit on $\eta$ is relaxed. However, with more precise
measurement on the relevant CKM matrix elements in future, we can
determine them more accurately.

In conclusion, the geometry restriction for Wolfenstein's
parameters is worked out. And, a positive $\rho$ is asserted by the
geometry postulation. Besides, we find $0.18 <\eta< 0.54$ and
$0.048<\rho< 0.140$ are permitted by the present data. These
results can be put to the more precisely tests in near future.

\newpage
\begin{figure}[htb]
\mbox{}
\vskip 7in\relax\noindent\hskip -1 in\relax
\includegraphics{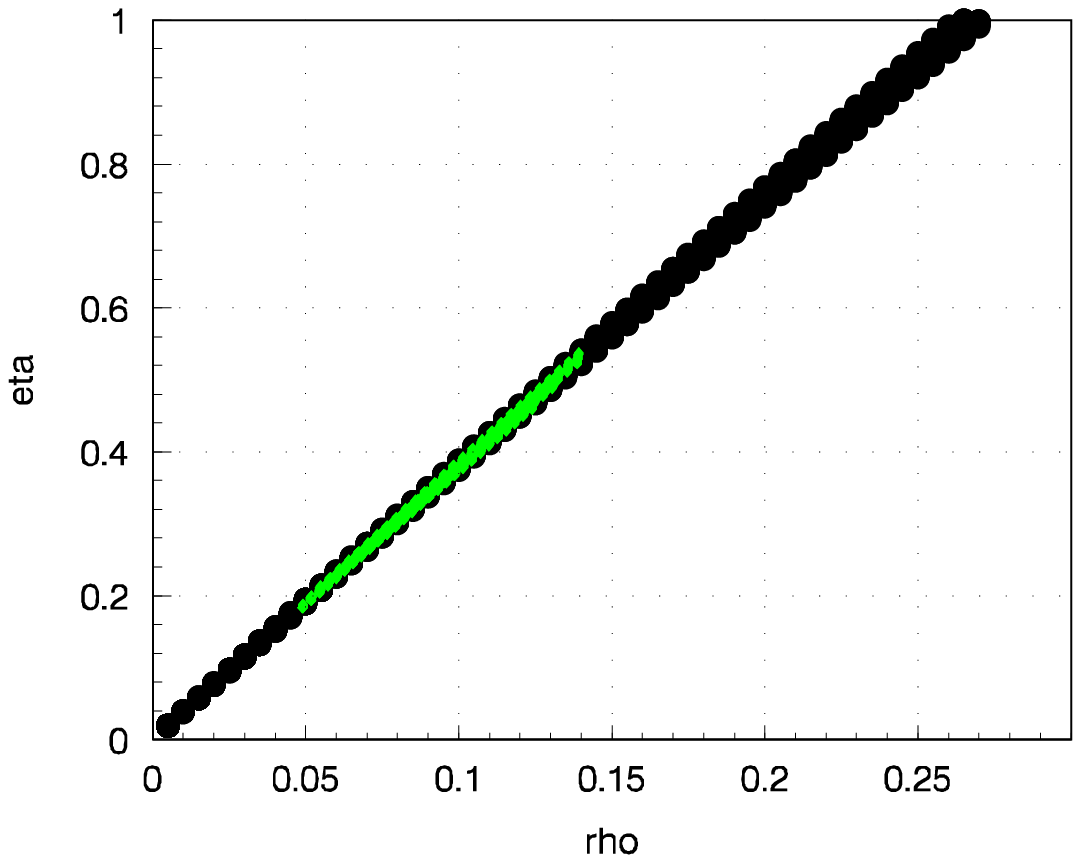}
\caption{The dependence of $\eta$ on $\rho$ based on Eq.(9) and
the permitted ranges for them by the present data. Here, the errors
of the inputs have been considered.}
\end{figure}

\end{document}